# Adaptive physics-informed neural networks for dynamic thermo-mechanical coupling problems in large-size-ratio functionally graded materials

Lin Qiu [a], Yanjie Wang [b], Tian He [a], Yan Gu [c], Fajie Wang [a,*]

[a] College of Mechanical and Electrical Engineering, National Engineering Research Center for Intelligent Electrical Vehicle Power System, Qingdao University, Qingdao 266071, PR China

[b] National Key Laboratory of Science and Technology on Multispectral Information Processing, School of Artificial Intelligence and Automation, Huazhong University of Science and Technology, Wuhan 430072, PR China

[c] School of Mathematics and Statistics, Qingdao University, Qingdao 266071, PR China

**Abstract**

In this paper, we present the adaptive physics-informed neural networks (PINNs) for resolving three dimensional (3D) dynamic thermo-mechanical coupling problems in large-size-ratio functionally graded materials (FGMs). The physical laws described by coupled governing equations and the constraints imposed by the initial and boundary conditions are leveraged to form the loss function of PINNs by means of the automatic differentiation algorithm, and an adaptive loss balancing scheme is introduced to improve the performance of PINNs. The adaptive PINNs are meshfree and trained on batches of randomly sampled collocation points, which is the key feature and superiority of the approach, since mesh-based methods will encounter difficulties in solving problems with large size ratios. The developed methodology is tested for several 3D thermo-mechanical coupling problems in large-size-ratio FGMs, and the numerical results demonstrate that the adaptive PINNs are effective and reliable for dealing with coupled problems in coating structures with large size ratios up to $10^9$, as well as complex large-size-ratio geometries such as the electrostatic comb, the airplane and the submarine.

*Keywords:* Physics-informed neural networks; Dynamic thermo-mechanical coupling; Functionally graded materials; Large size ratio; Loss balancing

[*] Corresponding author.
*E-mail address:* wfj88@qdu.edu.cn (F.J. Wang).

## 1. Introduction

Functionally graded materials (FGMs) [1, 2] are an innovative class of materials, whose properties vary continuously or nearly continuously with changes in dimension. Owing to their excellent properties, FGMs have demonstrated a wide range of practical and potential applications as materials and devices [3]. A considerable advantage of FGM lies in its ability to be applied in extreme environments (e.g., strong aerodynamic/impact, high temperature gradient), including aerospace engineering (e.g., airplane body, rocket engine component), microelectronics and energy (e.g., thermal barrier coatings) [2, 4]. Therefore, the thermodynamic behavior analysis of FGMs has important theoretical significance and engineering value [5-7].

Due to the irregular geometries, complex properties and harsh working conditions of materials or components, it is difficult to characterize the coupled mechanical behavior of FGMs by using analytical methods. Therefore, some numerical approaches such as the finite element method (FEM) [8, 9], the boundary element method (BEM) [10] and meshfree methods [11-14] are resorted to solve such problems. The current numerical schemes are faced with severe challenges in accurately handling problems defined in large-size-ratio FGMs or FGMs consisting of the size-disparity components, such as coating structures, aircraft models, etc. A vast number of meshes need to be generated to avoid the appearance of deformed elements when the popular FEM is employed for numerical simulation of large-size-ratio structures, which leads to a sharp increase in computing memory and time, and also has adverse effect on the calculation accuracy and reliability of the FEM [15, 16]. Compared with the FEM, the BEM shows certain advantages in addressing such problems owing to its only boundary meshing and semi-analytical feature [17]. However, the BEM has to pay the cost of calculating troublesome singular and near-singular integrals, and it is limited to issues for which the fundamental solutions can be determined [18, 19]. Some meshfree schemes are also developed and utilized to simulate the mechanical behavior of large-size-ratio structures, including the collocation Trefftz method [20], the singular boundary method [21, 22], etc. In addition, the deep learning approach could be viewed as an alternative to above mentioned methodologies for mechanical analysis of large-size-ratio FGMs.

Recently, physics-informed neural networks (PINNs) [23], conceived as deep learning algorithms, have witnessed a research boom. PINNs have shown great potential and have been successfully applied in solving engineering and scientific problems related to mechanics (e.g., fluid mechanics [24, 25], fracture mechanics [26] and thermodynamics [27]), materials [28, 29], biomedicine [30], energy [31, 32], etc. Subsequently, many variants of PINNs are developed to achieve extensive applications, such as conservative PINNs [33], fractional PINNs [34], Bayesian PINNs [35], and variational PINNs [36]. Moreover, some schemes like multi-GPU computing [37], self-adaptive approach [38] and importance sampling method [39] are designed to improve the performance of PINNs family. Inspired by the latest advance of PINNs, this study documents the first attempt to employ the adaptive PINNs for three dimensional (3D) dynamic thermo-mechanical coupling analysis of large-size-ratio FGMs. We summarize the advantages of PINNs for resolving such problems as follows. (i) Compared with other numerical approaches, PINNs are meshfree and have few restrictions on the distribution of data points, enabling PINNs to readily handle high-dimensional problems with complex large-size-ratio geometries, without involving mesh generation, numerical integration calculation, fundamental solution, and singularity and near-singularity elimination. (ii) Embedding the physical laws described by the governing equations, PINNs are capable of achieving better performance with a small number of data points in contrast to conventional deep learning methods. (iii) Multi-field coupling and material property variations of FGMs are conveniently embodied in the theory and procedure of PINNs, which can be programmed by utilizing existing optimization and deep learning toolboxes.

In our PINNs model, four neural networks are defined to approximate the displacement components in the mechanical field and temperature function in the thermal field, respectively. And then, we combine these networks together and train them via minimizing a composite loss function based on the physical laws characterized by coupled partial differential equations (PDEs) and the constraints given by the initial and boundary values. We assign a weight parameter to each loss item, and further develop the adaptive method proposed in Ref. [40] to balance the interactions among all loss terms, so that the weights are automatically adjusted during model training. Several benchmark examples are presented to examine the performance of the adaptive PINNs, including the large-size-ratio coating structures in Case 1, the electrostatic comb-shaped FGMs in Case 2, the airplane-shaped and the submarine-shaped FGMs in Case 3. It is noted that PINNs are effectively implemented via using MATLAB with optimization and deep learning toolboxes.

The rest of the paper is structured as follows. Section 2 introduces the mathematical model of the 3D dynamic thermo-mechanical coupling problems in FGMs. In Section 3, we provide the formulation and architecture of PINNs for solving the dynamic coupled problems, and present an adaptive loss balancing method for PINNs. In Section 4, the performance of the developed PINNs is verified through some representative examples. Finally, some concluding remarks are summarized in Section 5.

## 2. Problem statement

We assume that $\Omega$ is a 3D space domain bounded by a surface $\partial\Omega = \Gamma$, where $\Gamma = \Gamma_u \bigcup \Gamma_p = \Gamma_T \bigcup \Gamma_q$ and $\Gamma_u \bigcap \Gamma_p = \Gamma_T \bigcap \Gamma_q = \varnothing$. Considering the 3D dynamic thermo-mechanical coupling analysis of FGMs in $\Omega$, the temperature function $T(\boldsymbol{x},t)$ and displacement functions $u_i(\boldsymbol{x},t), i=1,2,3$ satisfy the following governing equations,

$$\begin{aligned}&\rho(\boldsymbol{x})\ddot{u}_i(\boldsymbol{x},t)+\beta(\boldsymbol{x})T_{,i}(\boldsymbol{x},t)-\left[\lambda(\boldsymbol{x})+\mu(\boldsymbol{x})\right]u_{j,ij}(\boldsymbol{x},t)\\&-\mu(\boldsymbol{x})u_{i,jj}(\boldsymbol{x},t)= f_i(\boldsymbol{x},t),\ \ \boldsymbol{x}\in\Omega,\ \ t\in[t_0,t_f],\end{aligned} \tag{1}$$

$$\rho(\boldsymbol{x})c(\boldsymbol{x})\dot{T}(\boldsymbol{x},t)+\beta(\boldsymbol{x})T_0\dot{u}_{i,i}(\boldsymbol{x},t)-\left[\kappa(\boldsymbol{x})T_{,i}(\boldsymbol{x},t)\right]_{,i}=s(\boldsymbol{x},t),\ \ \boldsymbol{x}\in\Omega,\ \ t\in[t_0,t_f], \tag{2}$$

where $\boldsymbol{x}=\left(x_1,x_2,x_3\right)$ is the spatial coordinate, $\rho(\boldsymbol{x})$ represents the density, $\kappa(\boldsymbol{x})$ stands for the thermal conductivity, and $c(\boldsymbol{x})$ denotes the specific heat. $f_i(\boldsymbol{x},t)$ and $s(\boldsymbol{x},t)$ indicate the known body forces and heat source, respectively. $T_0$ represents reference temperature. $t_0$ and $t_f$ are the initial and final time, respectively. In addition,

$$\lambda(\boldsymbol{x})=\frac{\nu}{(1+\nu)(1-2\nu)}E(\boldsymbol{x}),\ \ \mu(\boldsymbol{x})=\frac{1}{2(1+\nu)}E(\boldsymbol{x}),\ \ \beta(\boldsymbol{x})=\frac{\alpha}{1-2\nu}E(\boldsymbol{x}), \tag{3}$$

where $\nu$, $\alpha$ and $E(\boldsymbol{x})$ are Poisson's ratio, thermal expansion coefficient and elasticity modulus, respectively.

The governing equations (1) and (2) are subject to the coupled initial condition,

$$\begin{aligned}&u_i(\boldsymbol{x},t_0)=\hat{u}_i(\boldsymbol{x}),\ \ \boldsymbol{x}\in\Omega,\\&\dot{u}_i(\boldsymbol{x},t_0)=\hat{v}_i(\boldsymbol{x}),\ \ \boldsymbol{x}\in\Omega,\\&T(\boldsymbol{x},t_0)=\hat{T}(\boldsymbol{x}),\ \ \boldsymbol{x}\in\Omega,\end{aligned} \tag{4}$$

Dirichlet boundary conditions,

$$\begin{aligned}&u_i(\boldsymbol{x},t)=\bar{u}_i(\boldsymbol{x},t),\ \ \boldsymbol{x}\in\Gamma_u,\ \ t\in[t_0,t_f],\\&T(\boldsymbol{x},t)=\bar{T}(\boldsymbol{x},t),\ \ \boldsymbol{x}\in\Gamma_T,\ \ t\in[t_0,t_f],\end{aligned} \tag{5}$$

and Neumann boundary conditions,

$$\begin{aligned}&p_i(\boldsymbol{x},t)=\sigma_{ij}(\boldsymbol{x},t)n_j=\bar{p}_i(\boldsymbol{x},t),\ \ \boldsymbol{x}\in\Gamma_p,\ \ t\in[t_0,t_f],\\&q(\boldsymbol{x},t)=-\kappa(\boldsymbol{x})T_{,j}(\boldsymbol{x},t)n_j=\bar{q}(\boldsymbol{x},t),\ \ \boldsymbol{x}\in\Gamma_q,\ \ t\in[t_0,t_f],\end{aligned} \tag{6}$$

where $\hat{u}_i(\boldsymbol{x})$, $\hat{v}_i(\boldsymbol{x})$ and $\hat{T}(\boldsymbol{x})$ are the known values of displacement, velocity and temperature at $t_0$, respectively. $\bar{u}_i(\boldsymbol{x},t)$, $\bar{p}_i(\boldsymbol{x},t)$, $\bar{T}(\boldsymbol{x},t)$ and $\bar{q}(\boldsymbol{x},t)$ represent the given values of displacement, traction, temperature and normal heat flux specified on the corresponding boundary, respectively. $n_j$ indicates directional cosine of the outward unit normal vector. The stresses $\sigma_{ij}(\boldsymbol{x},t)$ are associated with the strains $\varepsilon_{ij}(\boldsymbol{x},t)$ and temperature $T(\boldsymbol{x},t)$ by using the following physical equation,

$$\sigma_{ij}(\boldsymbol{x},t)=\lambda(\boldsymbol{x})\varepsilon_{kk}(\boldsymbol{x},t)\delta_{ij}+2\mu(\boldsymbol{x})\varepsilon_{ij}(\boldsymbol{x},t)-\beta(\boldsymbol{x})T(\boldsymbol{x},t)\delta_{ij}, \tag{7}$$

in which $\delta_{ij}$ is Kronecker delta. The relationship between the strains $\varepsilon_{ij}(\boldsymbol{x},t)$ and displacements $u_i(\boldsymbol{x},t)$ can be established via utilizing the geometric equation,

$$\varepsilon_{ij}(\boldsymbol{x},t)=\frac{1}{2}\left[u_{i,j}(\boldsymbol{x},t)+u_{j,i}(\boldsymbol{x},t)\right]. \tag{8}$$

The general expressions for material property variations of FGMs, such as quadratic, exponential and trigonometric functions, are presented as follows [10],

(i) Exponential variation

$$\phi(\boldsymbol{x})=\phi_0\Pi_{i=1}^{d}\left(p_i+q_ix_i\right)^2. \tag{9}$$

(ii) Trigonometric variation

$$\phi(\boldsymbol{x})=\phi_0\Pi_{i=1}^{d}\left(p_i\mathrm{e}^{\alpha_i x_i}\right)^2. \tag{10}$$

(iii) Quadratic variation

$$\phi(\boldsymbol{x})=\phi_0\Pi_{i=1}^{d}\left[p_i\cos(\alpha_ix_i)+q_i\sin(\alpha_ix_i)\right]^2. \tag{11}$$

where $d$ is the spatial dimension, $\phi_0$, $p_i$, $q_i$ and $\alpha_i$ are constants. In this study, we will deal with 3D thermo-mechanical coupling problems in FGMs, where material property variations are quadratic, exponential, trigonometric or a combination of these functions.

## 3. Methodology

### *3.1. Formulation and architecture of PINNs*

In accordance with the framework of PINNs proposed in Ref. [23], the solution of the PDEs system can be approximated by a fully-connected network. To handle dynamic thermo-mechanical coupling problems described in Eqs. (1)-(2) and Eqs. (4)-(6), we define four neural networks to separately approximate the target solutions of the displacement components in the mechanical field and temperature function in the thermal field. Each network consists of multiple hidden layers, and takes coordinates $(\boldsymbol{x},t)$ and trial solution as initial inputs and final output, respectively. The inputs $\boldsymbol{\varsigma}=\left[\varsigma_1,\varsigma_2,\cdots,\varsigma_m\right]$ and outputs $\boldsymbol{\xi}=\left[\xi_1,\xi_2,\cdots,\xi_n\right]$ of each hidden layer are transmitted through the fully-connected network as follows,

$$\xi_n=\sigma\left(W_{m,n}\varsigma_m+b_n\right) \text{ or } \boldsymbol{\xi}=\sigma\left(\boldsymbol{W}\boldsymbol{\varsigma}+\boldsymbol{b}\right), \tag{12}$$

where $\sigma$ is the activation function such as Sigmoid, Tanh, Gaussian, Swish, Arctan, Mish, Softplus and ReLU. $\boldsymbol{W}$ and $\boldsymbol{b}$ respectively stand for the trainable weights and biases. The parameters of PINNs are trained to make the trial solutions, labeled $u_1\left(\boldsymbol{x},t;\boldsymbol{\theta}^{(1)}\right)$, $u_2\left(\boldsymbol{x},t;\boldsymbol{\theta}^{(2)}\right)$, $u_3\left(\boldsymbol{x},t;\boldsymbol{\theta}^{(3)}\right)$ and $T\left(\boldsymbol{x},t;\boldsymbol{\theta}^{(4)}\right)$ where $\boldsymbol{\theta}^{(i)}=\left(\boldsymbol{W}^{(i)},\boldsymbol{b}^{(i)}\right), i=1,2,3,4$, satisfy the coupled PDEs, the corresponding boundary and initial constraints. In our method, we couple four networks together and train them by minimizing the following composite loss function,

$$\begin{aligned}
L\left(\boldsymbol{\theta}^{(1)};\boldsymbol{\theta}^{(2)};\boldsymbol{\theta}^{(3)};\boldsymbol{\theta}^{(4)};N\right)=&\sum_{i=1}^{4}\lambda_{PDE}^{(i)}L_{PDE}^{(i)}\left(\boldsymbol{\theta}^{(1)};\boldsymbol{\theta}^{(2)};\boldsymbol{\theta}^{(3)};\boldsymbol{\theta}^{(4)};N_{PDE}\right)\\
&+\sum_{i=1}^{4}\lambda_{IC}^{(i)}L_{IC}^{(i)}\left(\boldsymbol{\theta}^{(i)};N_{IC}\right)+\sum_{i=1}^{3}\lambda_{IC'}^{(i)}L_{IC'}^{(i)}\left(\boldsymbol{\theta}^{(i)};N_{IC'}\right)\\
&+\sum_{i=1}^{3}\lambda_{NBC}^{(i)}L_{NBC}^{(i)}\left(\boldsymbol{\theta}^{(1)};\boldsymbol{\theta}^{(2)};\boldsymbol{\theta}^{(3)};\boldsymbol{\theta}^{(4)};N_{NBC}\right)+\lambda_{NBC}^{(4)}L_{NBC}^{(4)}\left(\boldsymbol{\theta}^{(4)};N_{NBC}\right)\\
&+\sum_{i=1}^{4}\lambda_{DBC}^{(i)}L_{DBC}^{(i)}\left(\boldsymbol{\theta}^{(i)};N_{DBC}\right),
\end{aligned} \tag{13}$$

where $\lambda_{PDE}^{(i)}$, $\lambda_{IC}^{(i)}$, $\lambda_{IC'}^{(i)}$, $\lambda_{NBC}^{(i)}$ and $\lambda_{DBC}^{(i)}$ represent the loss weight parameters. The PDEs residuals are defined,

$$\begin{cases}
\Re^{(i)}\left(\boldsymbol{x},t;\boldsymbol{\theta}^{(1)};\boldsymbol{\theta}^{(2)};\boldsymbol{\theta}^{(3)};\boldsymbol{\theta}^{(4)}\right)=\rho(\boldsymbol{x})\ddot{u}_i\left(\boldsymbol{x},t;\boldsymbol{\theta}^{(i)}\right)+\beta(\boldsymbol{x})T_{,i}\left(\boldsymbol{x},t;\boldsymbol{\theta}^{(4)}\right)\\
\qquad\qquad -\left[\lambda(\boldsymbol{x})+\mu(\boldsymbol{x})\right]u_{j,ij}\left(\boldsymbol{x},t;\boldsymbol{\theta}^{(j)}\right)\\
\qquad\qquad -\mu(\boldsymbol{x})u_{i,jj}\left(\boldsymbol{x},t;\boldsymbol{\theta}^{(i)}\right)-f_i(\boldsymbol{x},t),\ i=1,2,3,\\
\Re^{(4)}\left(\boldsymbol{x},t;\boldsymbol{\theta}^{(1)};\boldsymbol{\theta}^{(2)};\boldsymbol{\theta}^{(3)};\boldsymbol{\theta}^{(4)}\right)=\rho(\boldsymbol{x})c(\boldsymbol{x})\dot{T}\left(\boldsymbol{x},t;\boldsymbol{\theta}^{(4)}\right)+\beta(\boldsymbol{x})T_0\dot{u}_{i,i}\left(\boldsymbol{x},t;\boldsymbol{\theta}^{(i)}\right)\\
\qquad\qquad -\left[\kappa(\boldsymbol{x})T_{,i}\left(\boldsymbol{x},t;\boldsymbol{\theta}^{(4)}\right)\right]_{,i}-s(\boldsymbol{x},t),
\end{cases} \tag{14}$$

and then, the loss terms are described below,

$$L_{PDE}^{(i)}\left(\boldsymbol{\theta}^{(1)};\boldsymbol{\theta}^{(2)};\boldsymbol{\theta}^{(3)};\boldsymbol{\theta}^{(4)};N_{PDE}\right)=\frac{1}{N_{PDE}}\sum_{k=1}^{N_{PDE}}\left|\Re^{(i)}\begin{pmatrix}\boldsymbol{x}_{PDE}^{k},t_{PDE}^{k};\boldsymbol{\theta}^{(1)};\\ \boldsymbol{\theta}^{(2)};\boldsymbol{\theta}^{(3)};\boldsymbol{\theta}^{(4)}\end{pmatrix}\right|^2,\ i=1,2,3,4, \tag{15}$$

$$L_{IC}^{(i)}\left(\boldsymbol{\theta}^{(i)};N_{IC}\right)=\begin{cases}\dfrac{1}{N_{IC}}\displaystyle\sum_{k=1}^{N_{IC}}\left|u_i\left(\boldsymbol{x}_{IC}^{k},0;\boldsymbol{\theta}^{(i)}\right)-\hat{u}_i\left(\boldsymbol{x}_{IC}^{k}\right)\right|^2,\ i=1,2,3,\\ \dfrac{1}{N_{IC}}\displaystyle\sum_{k=1}^{N_{IC}}\left|T\left(\boldsymbol{x}_{IC}^{k},0;\boldsymbol{\theta}^{(i)}\right)-\hat{T}\left(\boldsymbol{x}_{IC}^{k}\right)\right|^2,\ i=4,\end{cases} \tag{16}$$

$$L_{IC'}^{(i)}\left(\boldsymbol{\theta}^{(i)};N_{IC'}\right)=\frac{1}{N_{IC'}}\sum_{k=1}^{N_{IC'}}\left|\dot{u}_i\left(\boldsymbol{x}_{IC'}^k,0;\boldsymbol{\theta}^{(i)}\right)-\hat{v}_i\left(\boldsymbol{x}_{IC'}^k\right)\right|^2,\ i=1,2,3, \tag{17}$$

$$\begin{cases} L_{NBC}^{(i)}\begin{pmatrix}\boldsymbol{\theta}^{(1)};\boldsymbol{\theta}^{(2)};\boldsymbol{\theta}^{(3)};\\ \boldsymbol{\theta}^{(4)};N_{NBC}\end{pmatrix}=\frac{1}{N_{NBC}}\sum_{k=1}^{N_{NBC}}\left|\sigma_{ij}\begin{pmatrix}\boldsymbol{x}_{NBC}^k,t_{NBC}^k;\boldsymbol{\theta}^{(1)};\\ \boldsymbol{\theta}^{(2)};\boldsymbol{\theta}^{(3)};\boldsymbol{\theta}^{(4)}\end{pmatrix}n_j-\bar{p}_i\left(\boldsymbol{x}_{NBC}^k,t_{NBC}^k\right)\right|^2,\ i=1,2,3,\\ L_{NBC}^{(4)}\left(\boldsymbol{\theta}^{(4)};N_{NBC}\right)=\frac{1}{N_{NBC}}\sum_{k=1}^{N_{NBC}}\left|-\kappa\left(\boldsymbol{x}_{NBC}^k\right)T_{,j}\left(\boldsymbol{x}_{NBC}^k,t_{NBC}^k;\boldsymbol{\theta}^{(4)}\right)n_j-\bar{q}\left(\boldsymbol{x}_{NBC}^k,t_{NBC}^k\right)\right|^2,\end{cases} \tag{18}$$

$$L_{DBC}^{(i)}\left(\boldsymbol{\theta}^{(i)};N_{DBC}\right)=\begin{cases}\frac{1}{N_{DBC}}\sum_{k=1}^{N_{DBC}}\left|u_i\left(\boldsymbol{x}_{DBC}^k,t_{DBC}^k;\boldsymbol{\theta}^{(i)}\right)-\bar{u}_i\left(\boldsymbol{x}_{DBC}^k,t_{DBC}^k\right)\right|^2,\ i=1,2,3,\\ \frac{1}{N_{DBC}}\sum_{k=1}^{N_{DBC}}\left|T\left(\boldsymbol{x}_{DBC}^k,t_{DBC}^k;\boldsymbol{\theta}^{(i)}\right)-\bar{T}\left(\boldsymbol{x}_{DBC}^k,t_{DBC}^k\right)\right|^2,\ i=4.\end{cases} \tag{19}$$

In the above equations, $L_{PDE}^{(i)}$, $L_{IC}^{(i)}/L_{IC'}^{(i)}$, $L_{NBC}^{(i)}$ and $L_{DBC}^{(i)}$ denote the losses of the coupled PDEs, initial, Neumann and Dirichlet boundary conditions, respectively. $\left\{\boldsymbol{x}_{PDE}^k,t_{PDE}^k\right\}_{k=1}^{N_{PDE}}$ indicate the collocation points that are randomly sampled inside the computational domain for governing equations, and $\left\{\boldsymbol{x}_{IC}^k,0\right\}_{k=1}^{N_{IC}}/\left\{\boldsymbol{x}_{IC'}^k,0\right\}_{k=1}^{N_{IC'}}$, $\left\{\boldsymbol{x}_{NBC}^k,t_{NBC}^k\right\}_{k=1}^{N_{NBC}}$ and $\left\{\boldsymbol{x}_{DBC}^k,t_{DBC}^k\right\}_{k=1}^{N_{DBC}}$ stand for the initial, Neumann and Dirichlet boundary training points, respectively. $N_{PDE}$, $N_{IC}/N_{IC'}$, $N_{NBC}$ and $N_{DBC}$ specify the numbers of data points for different terms.

The architecture of PINNs framework for resolving dynamic coupled problems is demonstrated in Fig. 1, in which four neural networks are configured and used to approximate the displacement components and temperature respectively, and then the resulting loss terms are combined in a single loss function for network training. The derivatives of $u_i\left(\boldsymbol{x},t;\boldsymbol{\theta}^{(i)}\right),i=1,2,3$ and $T\left(\boldsymbol{x},t;\boldsymbol{\theta}^{(4)}\right)$ are computed via utilizing the automatic differentiation algorithm. The parameters of the fully-connected network are trained by using gradient-descent methods based on the back-propagation of the loss function. It is noted that MATLAB with optimization and deep learning toolboxes is employed to implement PINNs efficiently, in which various software libraries are designed for deep learning and some functions are provided to facilitate the training of the network. In particular, using the automatic differentiation function “dlgradient” built into the deep learning toolbox, we can directly and efficiently obtain the derivatives, including the time and high-order derivatives.

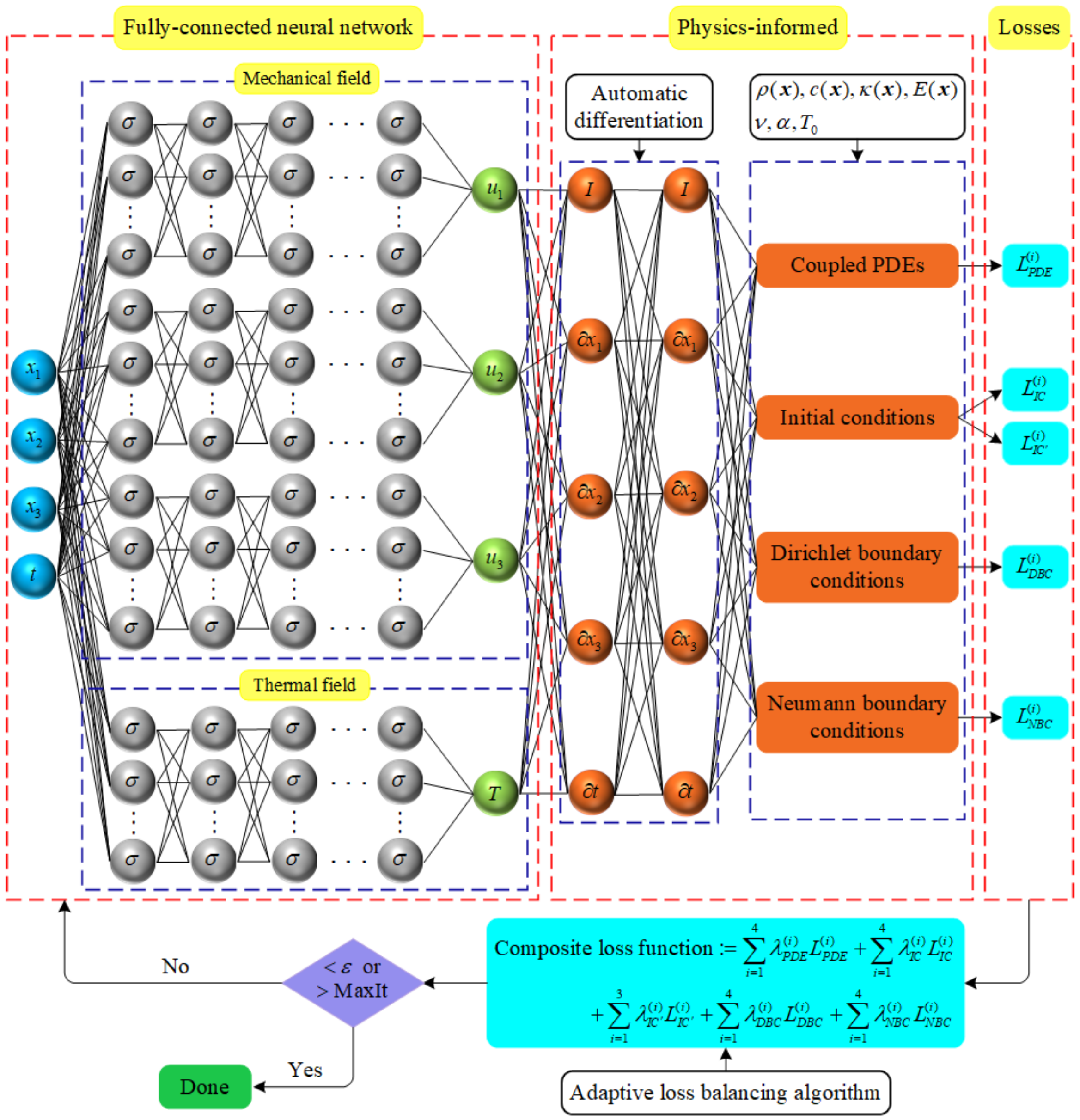

**Fig. 1.** Schematic diagram of PINNs.

### *3.2. Adaptive loss balancing algorithm for PINNs*

These loss weight parameters, i.e., $\lambda_{PDE}^{(i)}$, $\lambda_{IC}^{(i)}$, $\lambda_{IC'}^{(i)}$, $\lambda_{NBC}^{(i)}$ and $\lambda_{DBC}^{(i)}$, can be directly defined by the user or automatically adjusted during the training process. It is found that the weight setting of different loss terms is of great significance to improve the performance of PINNs, especially for problems involving high frequency or large-aspect-ratio geometries. To deal with dynamic thermo-mechanical coupling problems in large-size-ratio FGMs accurately and effectively, an adaptive algorithm extended from the work in Ref. [40] is utilized for balancing multiple loss terms. In this method, the weights are tuned automatically via using the back-propagated gradient statistics in the process of model training. The key step is to calculate the gradient magnitudes, and further obtain the following instantaneous values,

$$\lambda_{PDE}^{(i)} = \frac{\Theta}{\max\left\{\left|\nabla_{\left(\boldsymbol{\theta}^{(1)};\boldsymbol{\theta}^{(2)};\boldsymbol{\theta}^{(3)};\boldsymbol{\theta}^{(4)}\right)} L_{PDE}^{(i)}\left(\boldsymbol{\theta}^{(1)};\boldsymbol{\theta}^{(2)};\boldsymbol{\theta}^{(3)};\boldsymbol{\theta}^{(4)};N_{PDE}\right)\right|\right\}}, i = 1,2,3,4, \tag{20}$$

$$\lambda_{IC}^{(i)} = \frac{\Theta}{\text{mean}\left\{\left|\nabla_{\boldsymbol{\theta}^{(i)}} L_{IC}^{(i)}\left(\boldsymbol{\theta}^{(i)};N_{IC}\right)\right|\right\}}, i = 1,2,3,4, \tag{21}$$

$$\lambda_{IC'}^{(i)} = \frac{\Theta}{\text{mean}\left\{\left|\nabla_{\boldsymbol{\theta}^{(i)}} L_{IC'}^{(i)}\left(\boldsymbol{\theta}^{(i)};N_{IC'}\right)\right|\right\}}, i = 1,2,3, \tag{22}$$

$$\lambda_{NBC}^{(i)} = \frac{\Theta}{\text{mean}\left\{\left|\nabla_{\left(\boldsymbol{\theta}^{(1)};\boldsymbol{\theta}^{(2)};\boldsymbol{\theta}^{(3)};\boldsymbol{\theta}^{(4)}\right)} L_{NBC}^{(i)}\left(\boldsymbol{\theta}^{(1)};\boldsymbol{\theta}^{(2)};\boldsymbol{\theta}^{(3)};\boldsymbol{\theta}^{(4)};N_{NBC}\right)\right|\right\}}, i = 1,2,3, \tag{23}$$

$$\lambda_{NBC}^{(4)} = \frac{\Theta}{\text{mean}\left\{\left|\nabla_{\boldsymbol{\theta}^{(4)}} L_{NBC}^{(4)}\left(\boldsymbol{\theta}^{(4)};N_{NBC}\right)\right|\right\}}, \tag{24}$$

$$\lambda_{DBC}^{(i)} = \frac{\Theta}{\text{mean}\left\{\left|\nabla_{\boldsymbol{\theta}^{(i)}} L_{DBC}^{(i)}\left(\boldsymbol{\theta}^{(i)};N_{DBC}\right)\right|\right\}}, i = 1,2,3,4, \tag{25}$$

where $\Theta = \frac{1}{4}\left\langle \sum_{i=1}^{4} \max\left\{\left|\nabla_{\left(\boldsymbol{\theta}^{(1)};\boldsymbol{\theta}^{(2)};\boldsymbol{\theta}^{(3)};\boldsymbol{\theta}^{(4)}\right)} L_{PDE}^{(i)}\left(\boldsymbol{\theta}^{(1)};\boldsymbol{\theta}^{(2)};\boldsymbol{\theta}^{(3)};\boldsymbol{\theta}^{(4)};N_{PDE}\right)\right|\right\}\right\rangle$ denotes the mean of the sum of the maximum gradient vectors associated with the four PDEs. $\nabla_{(\bullet)}$ represents the gradient of the loss term with respect to the corresponding trainable parameters. Using this adaptive algorithm, the interactions among all loss terms including those corresponding to different PDEs, initial and boundary conditions can be properly balanced. Additionally, it should be pointed out that the updates defined in Eqs. (20)-(25) can be implemented at each iteration of the optimization loop, or at a user-specified frequency, e.g., every 20 training iterations.

## 4. Numerical examples and discussions

In this section, several benchmark numerical examples related to dynamic thermo-mechanical coupling problems in 3D FGMs will be solved by utilizing the adaptive PINNs. We consider cube-shaped FGMs and large-size-ratio coating structures in Case 1, where the material property variation is exponential, and the thermal conductivity, the density, the specific heat and the elasticity modulus are given as follows,

$$\begin{aligned} \kappa(\boldsymbol{x}) = \kappa_0 e^{0.4x_1+0.3x_2+0.2x_3}, \rho(\boldsymbol{x}) = \rho_0 e^{0.2x_1+0.2x_2+0.1x_3}, \\ c(\boldsymbol{x}) = c_0 e^{0.2x_1+0.1x_2+0.1x_3}, E(\boldsymbol{x}) = E_0 e^{0.3x_1+0.2x_2+0.1x_3}. \end{aligned} \tag{26}$$

An electrostatic comb-shaped FGM is examined in Case 2, where the material property variation is trigonometric, and

$$\begin{aligned}
\kappa(\boldsymbol{x}) &= \kappa_0 \left[\cos(4x_1)+\sin(4x_1)\right]^2 \left[\cos(3x_2)+\sin(3x_2)\right]^2 \left[\cos(2x_3)+\sin(2x_3)\right]^2, \\
\rho(\boldsymbol{x}) &= \rho_0 \left[\cos(4x_1)+\sin(4x_1)\right]\left[\cos(3x_2)+\sin(3x_2)\right]\left[\cos(2x_3)+\sin(2x_3)\right], \\
c(\boldsymbol{x}) &= c_0 \left[\cos(4x_1)+\sin(4x_1)\right]\left[\cos(3x_2)+\sin(3x_2)\right]\left[\cos(2x_3)+\sin(2x_3)\right], \\
E(\boldsymbol{x}) &= E_0 \left[\cos(3x_1)+\sin(3x_1)\right]\left[\cos(2x_2)+\sin(2x_2)\right]\left[\cos(x_3)+\sin(x_3)\right].
\end{aligned} \tag{27}$$

Case 3 presents the airplane-shaped and the submarine-shaped FGMs, where the material property variation is a combination of quadratic, exponential and trigonometric functions, and,

$$\begin{aligned}
\kappa(\boldsymbol{x}) &= \kappa_0 \left(0.5+0.004x_1\right)^2 e^{0.003x_2} \left[\cos(0.02x_3)+\sin(0.02x_3)\right]^2, \\
\rho(\boldsymbol{x}) &= \rho_0 \left(0.5+0.004x_1\right) e^{0.002x_2} \left[\cos(0.02x_3)+\sin(0.02x_3)\right], \\
c(\boldsymbol{x}) &= c_0 \left(0.5+0.004x_1\right) e^{0.001x_2} \left[\cos(0.02x_3)+\sin(0.02x_3)\right], \\
E(\boldsymbol{x}) &= E_0 \left(0.8+0.005x_1\right) e^{0.004x_2} \left[\cos(0.03x_3)+\sin(0.03x_3)\right].
\end{aligned} \tag{28}$$

In all examples, $\kappa_0 = 60\,\mathrm{W}/(\mathrm{m}\cdot{}^\circ\mathrm{C})$, $\rho_0 = 2555\,\mathrm{kg/m^3}$, $c_0 = 500\,\mathrm{J}/(\mathrm{kg}\cdot{}^\circ\mathrm{C})$, $E_0 = 1.25\times10^{11}\,\mathrm{Pa}$. The reference temperature $T_0$, Poisson's ratio $\nu$ and the thermal expansion coefficient $\alpha$ are $100\,^\circ\mathrm{C}$, 0.28 and 0.02 respectively. The adaptive loss balancing algorithm is implemented every 20 training iterations.

In all examples, we consider the following analytical solutions, for the mechanical field,

$$\begin{aligned}
u_1(\boldsymbol{x},t) &= 0.001\left[\sin(x_1+x_2+x_3)e^{-t}+x_2+x_3+0.0001\right], \\
u_2(\boldsymbol{x},t) &= 0.001\left[\sin(x_1+x_2+x_3)e^{-2t}+x_1+x_3+0.0001\right], \\
u_3(\boldsymbol{x},t) &= 0.001\left[\sin(x_1+x_2+x_3)e^{-3t}+x_1+x_2+0.0001\right],
\end{aligned} \tag{29}$$

and for the thermal field,

$$T(\boldsymbol{x},t) = \cos(x_1)\cos(x_2)\cos(x_3)\cos(t)+18x_1+18x_2+18x_3+200. \tag{30}$$

The body forces and heat source can be further derived with the given analytical solutions. In addition, for the purpose of illustrating the performance of the methodology, the following relative error and global error are used,

$$\text{Relative error} = \left|I_{exact}(k)-I_{numerical}(k)\right| / \left|I_{exact}(k)\right|, \tag{31}$$

$$\text{Global error} = \sqrt{\sum_{k=1}^{N_{total}}\left[I_{exact}(k)-I_{numerical}(k)\right]^2} \bigg/ \sqrt{\sum_{k=1}^{N_{total}}\left[I_{exact}(k)\right]^2}, \tag{32}$$

where $I_{exact}$ and $I_{numerical}$ represent the analytical and numerical solutions, respectively, and $N_{total}$ denotes the total number of test nodes.

### *4.1. Case 1: Thermo-mechanical coupling analysis in cube-shaped FGMs and large-size -ratio structures*

#### *4.1.1. Cube-shaped FGM*

We first consider a dynamic thermo-mechanical coupling problem defined in a cubic domain $\Omega = [0\text{m}, 1\text{m}] \times [0\text{m}, 1\text{m}] \times [0\text{m}, 1\text{m}]$ shown in Fig. 2(a) and $t \in [0\text{s}, 1\text{s}]$, and subjected to Dirichlet boundary conditions, where the values of temperature and displacement are known on all surfaces. To solve this problem using the adaptive PINNs, we arrange 729 collocation points (black nodes) inside the computational domain and 486 training points (red nodes) on the surface as displayed in Fig. 2(a), and the time step is specified as $\Delta t = 0.1\text{s}$. The network consisting of 4 fully-connected hidden layers with 15 neurons per layer and Swish function is trained via employing these data points and 2000 iteration steps. Relative errors of the stresses and heat fluxes including $\sigma_{11}$, $\sigma_{22}$, $\sigma_{33}$, $T_{,1}$, $T_{,2}$ and $T_{,3}$ on the cube surface at $t = 0.95\text{s}$ are presented in Figs. 3(a)-3(f), in which almost all relative errors of the stresses and heat fluxes are smaller than $1.90 \times 10^{-5}$ and $1.80 \times 10^{-3}$, respectively. Additionally, the achieved global errors of $\sigma_{11}$, $\sigma_{22}$, $\sigma_{33}$, $T_{,1}$, $T_{,2}$ and $T_{,3}$ are $5.45 \times 10^{-6}$, $5.21 \times 10^{-6}$, $4.69 \times 10^{-6}$, $2.70 \times 10^{-4}$, $2.79 \times 10^{-4}$ and $2.56 \times 10^{-4}$, respectively. These results strongly demonstrate the validity and feasibility of the developed PINNs for 3D dynamic thermo-mechanical coupling analysis of FGMs.

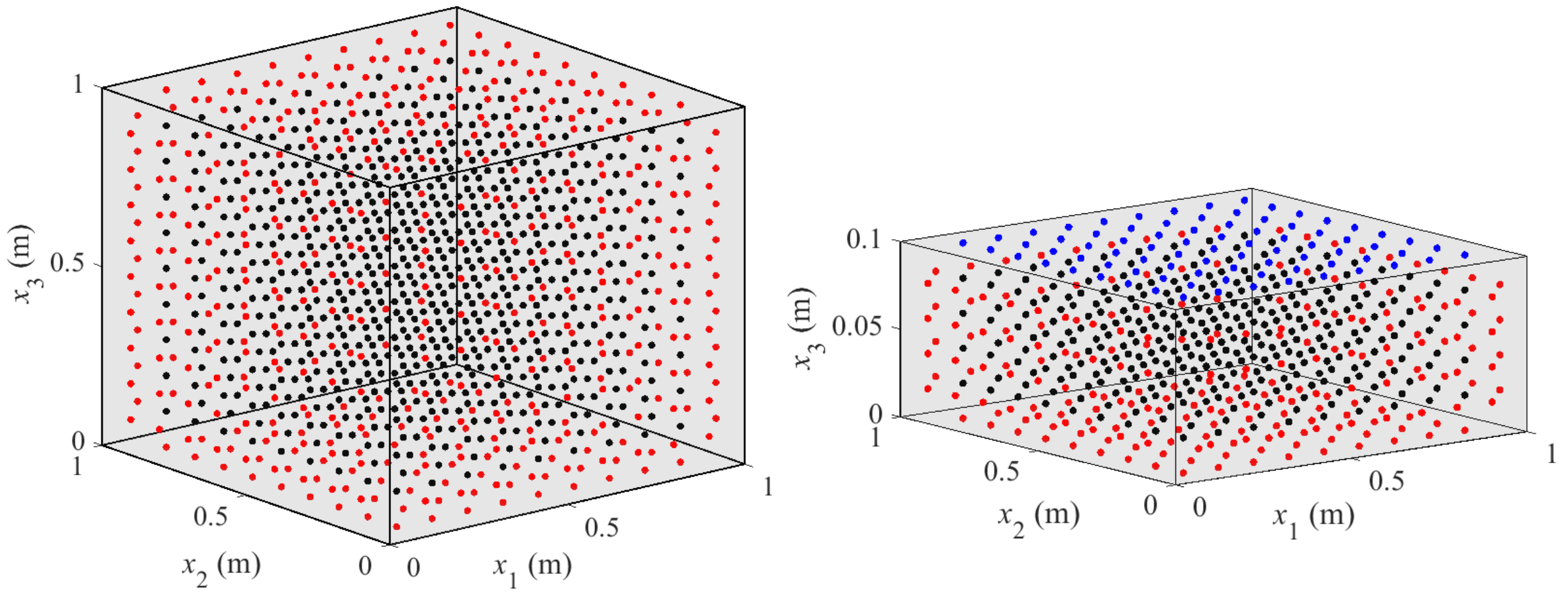

(a) (b)

**Fig. 2.** (a) Sketch of the cube and the distribution of collocation points (black nodes) and training points (red nodes), (b) sketch of the large-size-ratio coating structure and the distribution of collocation points (black nodes) and training points (red and blue nodes).

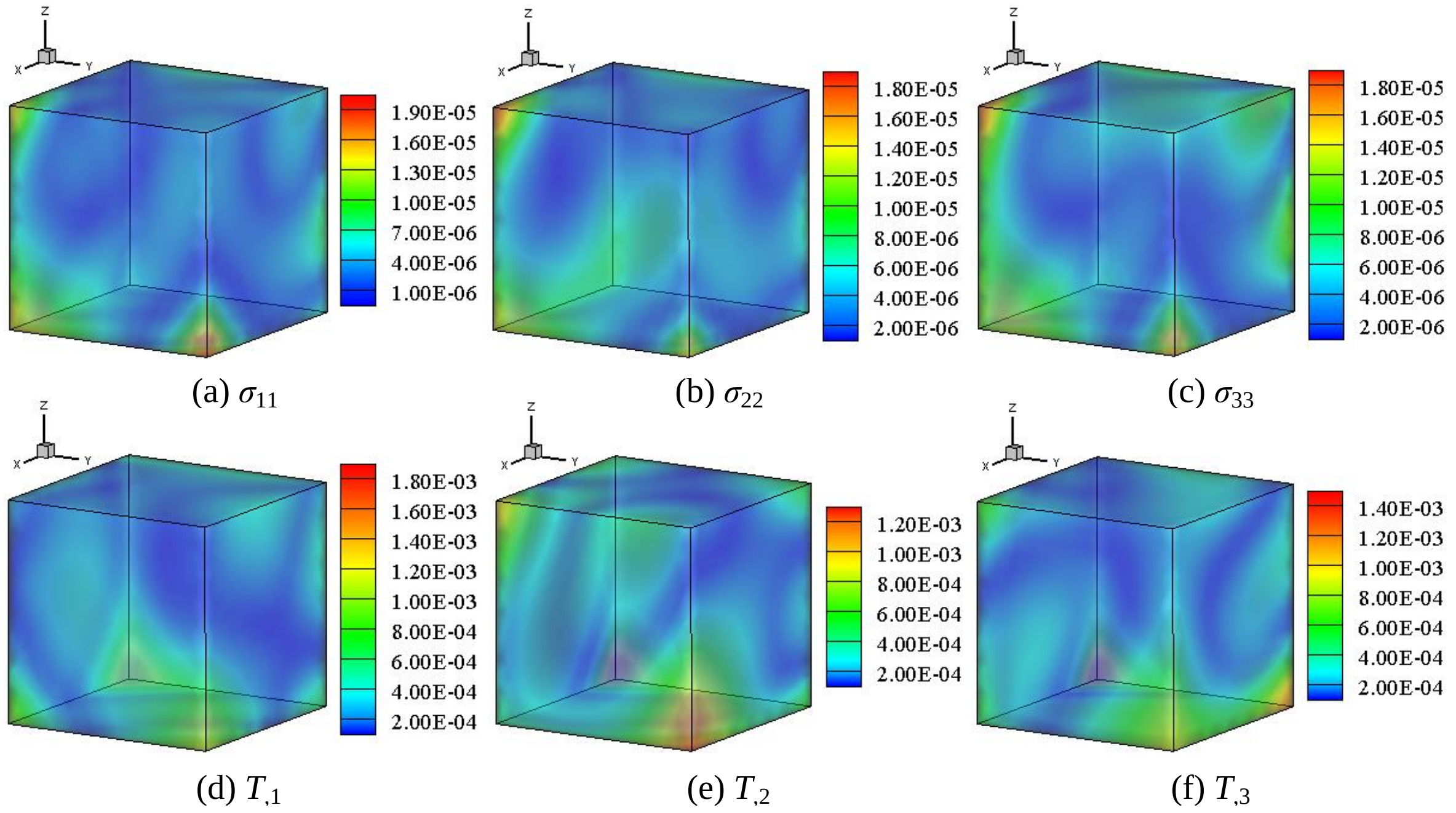

(d) $T_{,1}$ (e) $T_{,2}$ (f) $T_{,3}$

**Fig. 3.** Relative errors of stresses and heat fluxes on the cube surface at $t = 0.95\text{s}$ obtained by the adaptive PINNs.

**Table 1** Global errors of stresses and heat fluxes calculated by the adaptive PINNs with different network architectures.

| Network architecture | | Stress | | | Heat flux | | |
|---|---|---|---|---|---|---|---|
| Hidden layers | Neurons per layer | $\sigma_{11}$ | $\sigma_{22}$ | $\sigma_{33}$ | $T_{,1}$ | $T_{,2}$ | $T_{,3}$ |
| 2 | 5 | $2.34\times10^{-4}$ | $2.23\times10^{-4}$ | $2.39\times10^{-4}$ | $6.87\times10^{-3}$ | $9.25\times10^{-3}$ | $8.66\times10^{-3}$ |
| | 10 | $3.62\times10^{-5}$ | $3.66\times10^{-5}$ | $3.52\times10^{-5}$ | $1.35\times10^{-3}$ | $1.51\times10^{-3}$ | $1.36\times10^{-3}$ |
| | 15 | $2.70\times10^{-5}$ | $2.51\times10^{-5}$ | $2.78\times10^{-5}$ | $1.00\times10^{-3}$ | $1.03\times10^{-3}$ | $1.01\times10^{-3}$ |
| | 20 | $2.63\times10^{-5}$ | $2.88\times10^{-5}$ | $2.52\times10^{-5}$ | $1.13\times10^{-3}$ | $1.10\times10^{-3}$ | $1.08\times10^{-3}$ |
| 3 | 5 | $3.52\times10^{-5}$ | $3.40\times10^{-5}$ | $3.49\times10^{-5}$ | $1.41\times10^{-3}$ | $1.36\times10^{-3}$ | $1.40\times10^{-3}$ |
| | 10 | $3.00\times10^{-5}$ | $2.97\times10^{-5}$ | $3.01\times10^{-5}$ | $1.12\times10^{-3}$ | $1.13\times10^{-3}$ | $1.13\times10^{-3}$ |
| | 15 | $1.78\times10^{-5}$ | $1.81\times10^{-5}$ | $1.54\times10^{-5}$ | $7.85\times10^{-4}$ | $8.11\times10^{-4}$ | $8.39\times10^{-4}$ |
| | 20 | $1.79\times10^{-5}$ | $1.82\times10^{-5}$ | $1.86\times10^{-5}$ | $6.98\times10^{-4}$ | $8.08\times10^{-4}$ | $6.09\times10^{-4}$ |
| 4 | 5 | $2.07\times10^{-5}$ | $2.06\times10^{-5}$ | $2.05\times10^{-5}$ | $8.87\times10^{-4}$ | $7.63\times10^{-4}$ | $8.48\times10^{-4}$ |
| | 10 | $2.04\times10^{-5}$ | $1.95\times10^{-5}$ | $2.07\times10^{-5}$ | $7.95\times10^{-4}$ | $7.71\times10^{-4}$ | $7.90\times10^{-4}$ |
| | 15 | $1.35\times10^{-5}$ | $1.39\times10^{-5}$ | $1.41\times10^{-5}$ | $4.92\times10^{-4}$ | $5.86\times10^{-4}$ | $5.84\times10^{-4}$ |
| | 20 | $1.50\times10^{-5}$ | $1.54\times10^{-5}$ | $1.53\times10^{-5}$ | $5.92\times10^{-4}$ | $5.66\times10^{-4}$ | $6.01\times10^{-4}$ |

The network framework of the adaptive PINNs plays a crucial role in its performance. Generally, different network frameworks need to be configured for different application scenarios. Table 1 presents the global errors of stresses ($\sigma_{11}$, $\sigma_{22}$ and $\sigma_{33}$) and heat fluxes ($T_{,1}$, $T_{,2}$ and $T_{,3}$) attained via using diverse network architectures with Swish function and 2000 iteration steps. It is observed that the errors achieved by the adaptive PINNs are quite small, and show good robustness and stability. In general, the numerical accuracy will be improved to some extent, as the number of hidden layers or neurons per layer increases.

### *4.1.2. Large-size-ratio coating structure*

In this sub-section, we consider coating structures with different size ratios to verify the performance of the adaptive PINNs for coupled problems defined in large-size-ratio geometries. We select 324 collocation points (black nodes) inside the coating structure shown in Fig. 2(b) and 306 training points (red and blue nodes) on the surface, and set time step $\Delta t = 0.1\text{s}$ to train the network consisting of 3 fully-connected hidden layers with 15 neurons per layer and Swish function utilizing 2000 iteration steps. The upper surface of the coating is subject to Neumann boundary conditions (blue nodes), and the remaining surfaces are subject to Dirichlet boundary conditions (red nodes). Table 2 gives global errors of stresses and heat fluxes calculated by the adaptive PINNs, as the size-ratio of coating varies from 10 to $10^9$. The achieved global errors clearly demonstrate the accuracy and stability of the adaptive PINNs for dealing with thermo-mechanical coupling problems in FGMs with large-size-ratio coating structures. Notably, the same numbers of collocation and training points are employed for structures with different size ratios, which is found to have little effect on numerical accuracy of the adaptive PINNs. The traditional mesh-based approaches have to resize the meshes according to the thickness of the structure due to the requirement of element aspect-ratio. Once the thickness of the structure becomes quite small, these methods are somewhat limited by storage space and computing time. Therefore, one of the key advantages of PINNs for resolving large-size-ratio structures is that they could be efficiently trained with relatively small data points and achieve satisfactory accuracy.

In the following, we explore the performance of the adaptive PINNs in resolving the coupled problems with large-size-ratio geometries when employing different activation functions, including Sigmoid, Tanh, Swish, Softplus, Arctan and Mish. We consider a structure with a size-ratio of $10^5$ and train the network using the same data and settings as above. The global errors of stresses and heat fluxes obtained by the adaptive PINNs are presented in Table 3, in which all global errors of the stresses and heat fluxes are smaller than $3.35\times10^{-5}$ and $1.32\times10^{-3}$, respectively. The results show that the developed PINNs can obtain very high accuracy in all cases, indicating that its performance is relatively insensitive to the selection of activation function.

**Table 2** Global errors of stresses and heat fluxes attained by the adaptive PINNs.

| Size-ratio | Stress | | | Heat flux | | |
|---|---|---|---|---|---|---|
| | $\sigma_{11}$ | $\sigma_{22}$ | $\sigma_{33}$ | $T_{,1}$ | $T_{,2}$ | $T_{,3}$ |
| 10 | $4.94\times10^{-6}$ | $6.14\times10^{-6}$ | $3.93\times10^{-6}$ | $1.57\times10^{-4}$ | $1.86\times10^{-4}$ | $1.82\times10^{-4}$ |
| $10^2$ | $3.28\times10^{-6}$ | $3.49\times10^{-6}$ | $2.43\times10^{-6}$ | $3.83\times10^{-4}$ | $2.60\times10^{-4}$ | $1.90\times10^{-4}$ |
| $10^3$ | $4.65\times10^{-6}$ | $5.18\times10^{-6}$ | $4.84\times10^{-6}$ | $3.97\times10^{-4}$ | $3.56\times10^{-4}$ | $2.14\times10^{-5}$ |
| $10^4$ | $3.33\times10^{-6}$ | $3.12\times10^{-6}$ | $3.42\times10^{-6}$ | $2.04\times10^{-4}$ | $2.31\times10^{-4}$ | $7.12\times10^{-6}$ |
| $10^5$ | $1.88\times10^{-6}$ | $2.54\times10^{-6}$ | $1.75\times10^{-6}$ | $2.35\times10^{-4}$ | $2.40\times10^{-4}$ | $1.39\times10^{-5}$ |
| $10^6$ | $1.14\times10^{-6}$ | $1.10\times10^{-6}$ | $7.11\times10^{-7}$ | $8.89\times10^{-5}$ | $1.35\times10^{-4}$ | $7.54\times10^{-6}$ |
| $10^7$ | $3.15\times10^{-6}$ | $3.51\times10^{-6}$ | $2.24\times10^{-6}$ | $2.64\times10^{-4}$ | $3.23\times10^{-4}$ | $7.97\times10^{-6}$ |
| $10^8$ | $1.09\times10^{-6}$ | $1.35\times10^{-6}$ | $1.23\times10^{-6}$ | $1.43\times10^{-4}$ | $9.20\times10^{-5}$ | $4.36\times10^{-6}$ |
| $10^9$ | $1.67\times10^{-6}$ | $1.45\times10^{-6}$ | $1.46\times10^{-6}$ | $1.25\times10^{-4}$ | $2.01\times10^{-4}$ | $9.35\times10^{-6}$ |

**Table 3** Performance of the adaptive PINNs utilizing different activation functions.

| Activation function | Stress | | | Heat flux | | |
|---|---|---|---|---|---|---|
| | $\sigma_{11}$ | $\sigma_{22}$ | $\sigma_{33}$ | $T_{,1}$ | $T_{,2}$ | $T_{,3}$ |
| Sigmoid | $3.31\times10^{-6}$ | $3.46\times10^{-6}$ | $1.19\times10^{-6}$ | $3.46\times10^{-4}$ | $3.67\times10^{-4}$ | $2.48\times10^{-5}$ |
| Tanh | $2.06\times10^{-5}$ | $2.38\times10^{-5}$ | $3.35\times10^{-5}$ | $9.54\times10^{-4}$ | $9.28\times10^{-4}$ | $1.05\times10^{-3}$ |
| Swish | $1.88\times10^{-6}$ | $2.54\times10^{-6}$ | $1.75\times10^{-6}$ | $2.35\times10^{-4}$ | $2.40\times10^{-4}$ | $1.39\times10^{-5}$ |
| Softplus | $1.83\times10^{-6}$ | $1.32\times10^{-6}$ | $1.47\times10^{-6}$ | $5.56\times10^{-5}$ | $1.13\times10^{-4}$ | $8.56\times10^{-6}$ |
| Arctan | $2.35\times10^{-5}$ | $2.60\times10^{-5}$ | $3.12\times10^{-5}$ | $9.10\times10^{-4}$ | $1.32\times10^{-3}$ | $4.81\times10^{-4}$ |
| Mish | $1.37\times10^{-6}$ | $1.62\times10^{-6}$ | $1.38\times10^{-6}$ | $7.04\times10^{-5}$ | $7.00\times10^{-5}$ | $4.90\times10^{-6}$ |

### *4.2. Case 2: Thermo-mechanical coupling analysis in an electrostatic comb-shaped FGM*

In this example, we resolve a dynamic thermo-mechanical coupling problem in an electrostatic comb-shaped FGM given in Fig. 4(a) and $t\in[0\text{s},2\text{s}]$. The principal dimensions of the electrostatic comb are $6.44\times10^{-3}\text{m}$ in length, $2.32\times10^{-3}\text{m}$ in width, and $2.00\times10^{-4}\text{m}$ in height, and the size ratio of this structure is 32.20. It should be pointed out that the $x_3$-axis of the electrostatic comb is magnified by a factor of two compared to $x_1$- and $x_2$-axes, in order to provide a better reading experience. The 3D model is constrained by hybrid boundary conditions, where we specify Dirichlet boundary conditions on the left-half surface of the electrostatic comb ($1.16\times10^{-3}\text{m}\le x_2 \le 2.32\times10^{-3}\text{m}$) and Neumann boundary conditions on the remaining surfaces.

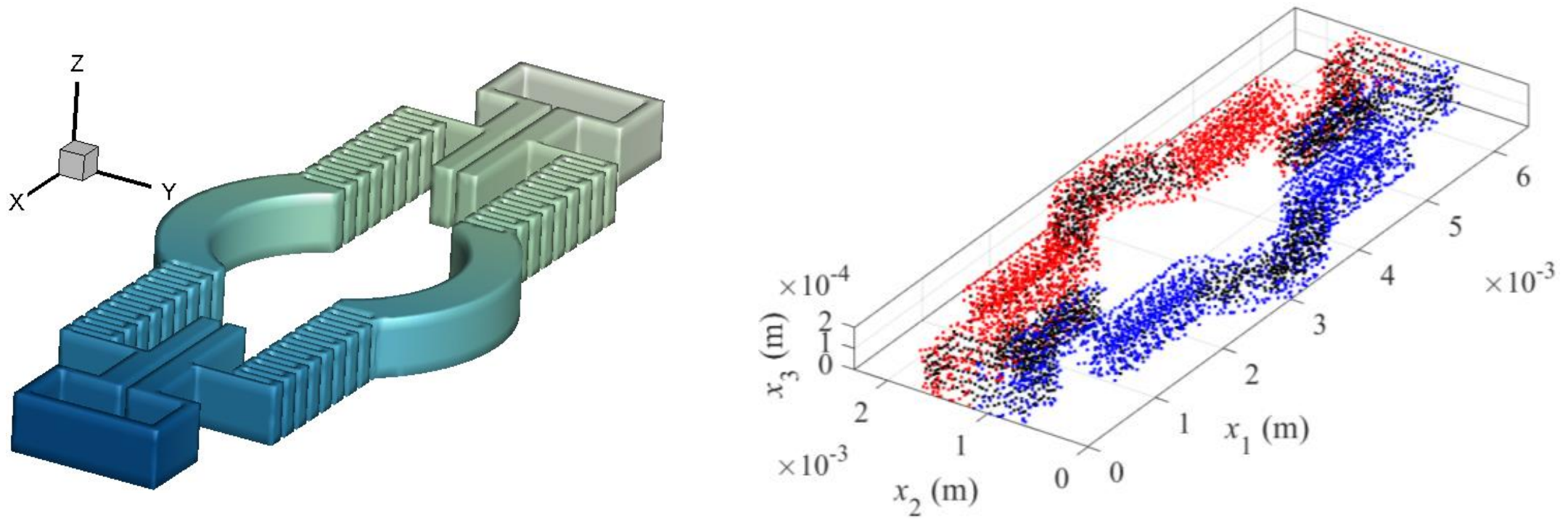

(a) (b)

**Fig. 4.** (a) Sketch of the electrostatic comb, and (b) the distribution of collocation points (black nodes) and training points (red and blue nodes) for the adaptive PINNs in space domain.

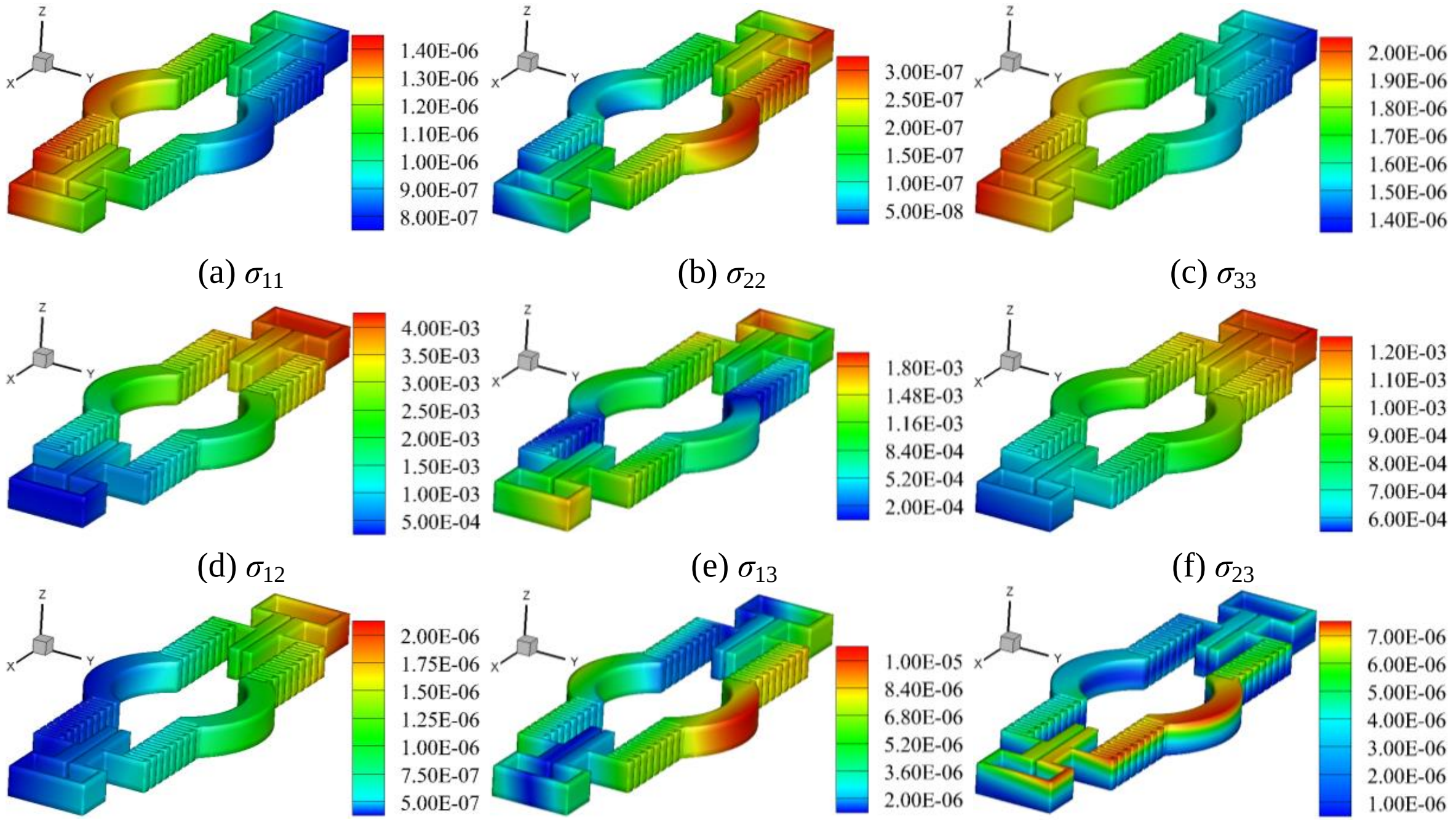

(a) $\sigma_{11}$ (b) $\sigma_{22}$ (c) $\sigma_{33}$

(d) $\sigma_{12}$ (e) $\sigma_{13}$ (f) $\sigma_{23}$

(g) $T_{,1}$ (h) $T_{,2}$ (i) $T_{,3}$

**Fig. 5.** Relative errors of stresses and heat fluxes on the electrostatic comb surface at $t = 1.9\text{s}$ calculated by the adaptive PINNs.

We distribute 2228 collocation points (black nodes in Fig. 4(b)) inside the computational domain and 4043 training points (red and blue nodes in Fig. 4(b)) along the surface, and set time step $\Delta t = 0.2\text{s}$ for the adaptive PINNs. The network composing of 3 fully-connected hidden layers with Swish function and 15 neurons per layer is trained by using these datasets and 2000 iteration steps. Figs. 5(a)-5(i) plot the relative error surfaces of the stresses and heat fluxes including $\sigma_{11}$, $\sigma_{22}$, $\sigma_{33}$, $\sigma_{12}$, $\sigma_{13}$, $\sigma_{23}$, $T_{,1}$, $T_{,2}$ and $T_{,3}$ at $t = 1.9\text{s}$, in which relative errors of the stresses or heat fluxes are smaller than $1.40\times10^{-6}$, $3.00\times10^{-7}$, $2.00\times10^{-6}$, $4.00\times10^{-3}$, $1.80\times10^{-3}$, $1.20\times10^{-3}$, $2.00\times10^{-6}$, $1.00\times10^{-5}$, $7.00\times10^{-6}$, respectively.

In addition, we calculate the global errors of $\sigma_{11}$, $\sigma_{22}$, $\sigma_{33}$, $T_{,1}$, $T_{,2}$ and $T_{,3}$ at different times, and present the results in Fig. 6. It is found that the numerical accuracy is quite high, demonstrating that the adaptive PINNs are effective for dynamic thermo-mechanical coupling problems in 3D FGMs with complex large-size-ratio geometries.

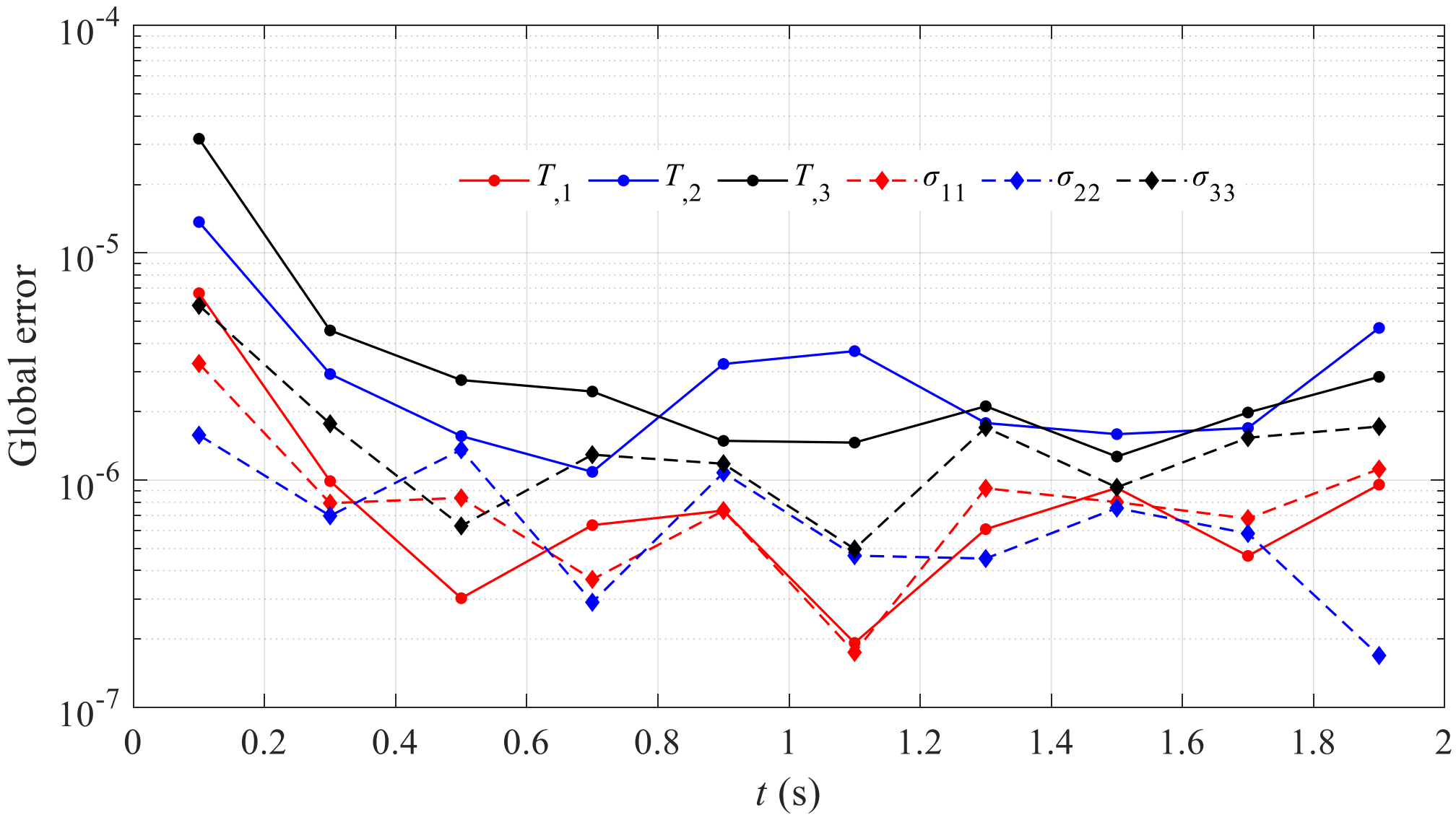

**Fig. 6.** Global errors of stresses and heat fluxes obtained by the adaptive PINNs at different times.

### *4.3. Case 3: Thermo-mechanical coupling analysis in the airplane-shaped and the submarine-shaped FGMs*

Finally, we simulate a dynamic coupled problem in the airplane-shaped and the submarine-shaped FGMs as shown in Figs. 7(a) and 8(a), respectively, and $t \in [0\text{s}, 1\text{s}]$. The principal dimension of the airplane is 7.66m × 2.14m × 10.00m, and the size ratios of the airplane fuselage and wings are 8.85 and 11.50, respectively. The principal dimension of the submarine is 2.08m × 3.23m × 25.00m, and the size ratio of this geometry is 12.00. Both models are subjected to a mixed-type boundary conditions, where Dirichlet boundary conditions are imposed on the left-half surface of the airplane ($3.83\text{m} \le x_1 \le 7.66\text{m}$) and the upper surface of the submarine ($1.20\text{m} \le x_2 \le 3.23\text{m}$), and Neumann boundary conditions are specified on the remaining surfaces of the airplane and the submarine, respectively.

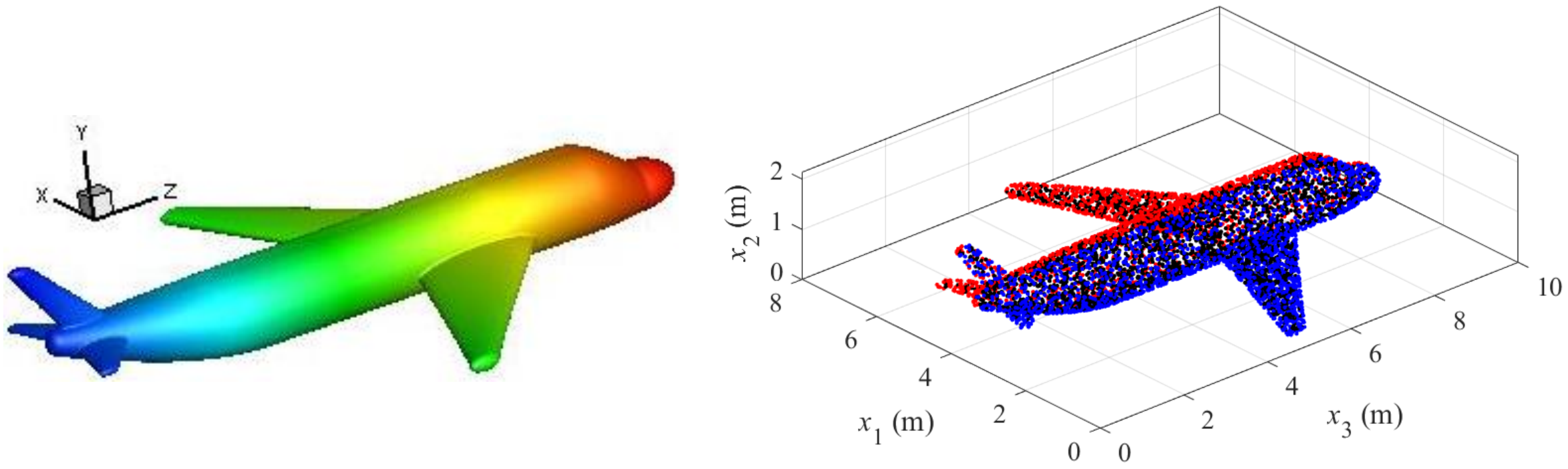

(a) (b)

**Fig. 7.** (a) Sketch of the airplane, and (b) the distribution of collocation points (black nodes) and training points (red and blue nodes) for the adaptive PINNs in space domain.

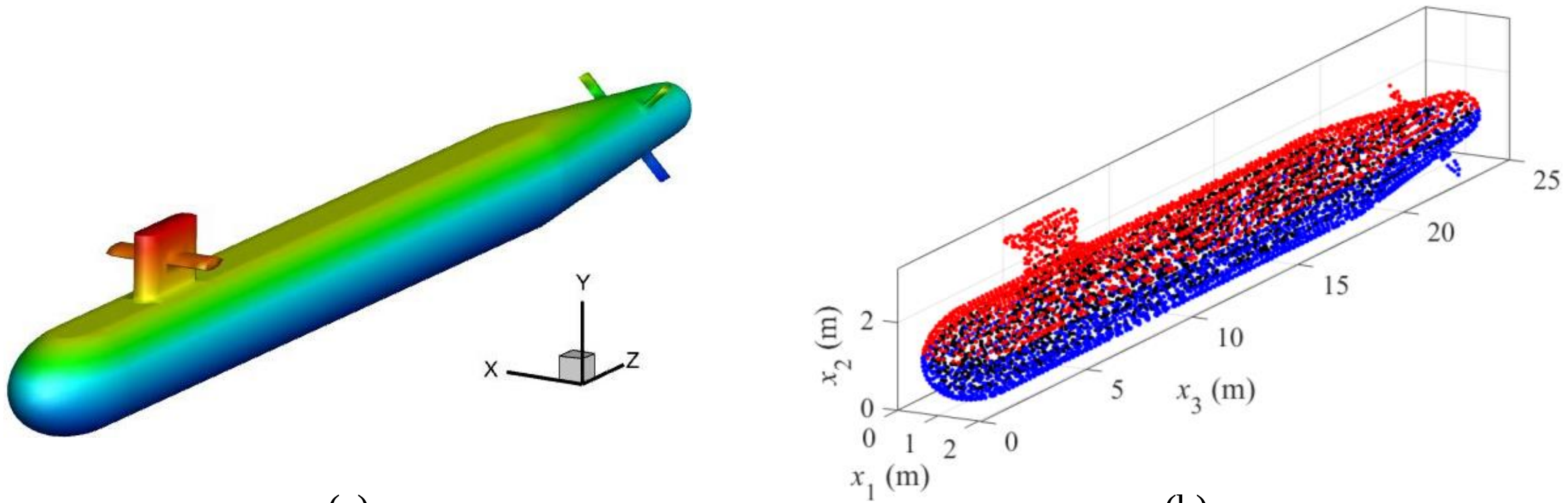

(a) (b)

**Fig. 8.** (a) Sketch of the submarine, and (b) the distribution of collocation points (black nodes) and training points (red and blue nodes) for the adaptive PINNs in space domain.

For the purpose of handling this problem with the adaptive PINNs, 2913 collocation points (black nodes) and 3705 training points (red and blue nodes) are respectively arranged inside the airplane domain and along the airplane surface, as displayed in Fig. 7(b), and time step $\Delta t$ is taken as $0.2\text{s}$. The network consisting of 7 fully-connected hidden layers with Mish function and 20 neurons per layer is trained by employing these datasets and 3000 iteration steps. Relative errors of the stresses and heat fluxes including $\sigma_{11}$, $\sigma_{22}$, $\sigma_{33}$, $T_{,1}$, $T_{,2}$ and $T_{,3}$ on the airplane surface at $t = 0.9\text{s}$ are presented in Figs. 9(a)-9(f), in which almost all relative errors of the stresses and heat fluxes are smaller than $1.00\times10^{-4}$ and $9.50\times10^{-3}$, respectively.

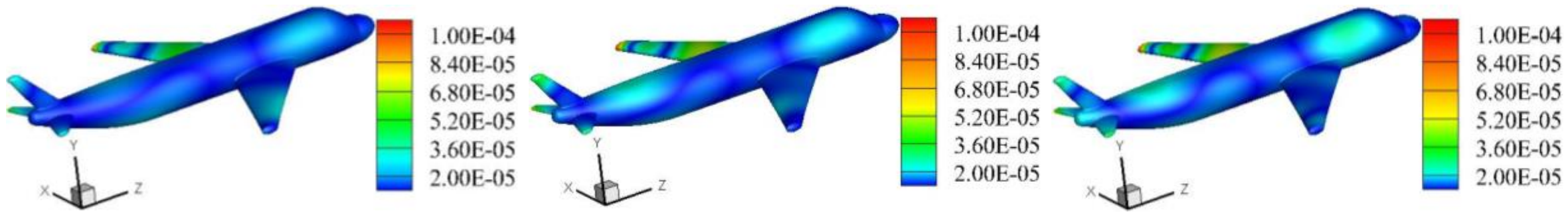

(a) $\sigma_{11}$ (b) $\sigma_{22}$ (c) $\sigma_{33}$

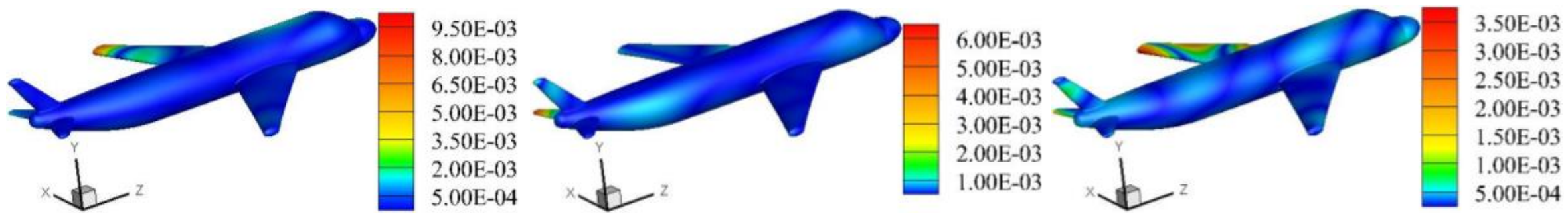

(d) $T_{,1}$ (e) $T_{,2}$ (f) $T_{,3}$

**Fig. 9.** Relative errors of stresses and heat fluxes on the airplane surface at $t = 0.9\text{s}$ calculated by the adaptive PINNs.

For the submarine-shaped FGM, we select 1834 collocation points (black nodes) inside the submarine domain and 4006 training points (red and blue nodes) along the submarine surface, as shown in Fig. 8(b), and set time step $\Delta t = 0.2\text{s}$ for the adaptive PINNs. Using these data points and 3000 iteration steps, we train the network composing of 10 fully-connected hidden layers with Mish function and 25 neurons per layer to solve the coupled problem. Figs. 10(a)-10(c) plot the relative error surfaces of the stresses at $t = 0.9\text{s}$, in which almost all relative errors of the stresses are smaller than $8.00\times10^{-4}$. The comparison of numerical and analytical heat flux distributions on the submarine surface at $t = 0.5\text{s}$ is given in Figs. 11(a)-11(f), and it is found that the numerical results are in good agreement with the analytical results. Additionally, Table 4 gives the global errors of $\sigma_{11}$, $\sigma_{22}$, $\sigma_{33}$, $T_{,1}$, $T_{,2}$ and $T_{,3}$ for the two cases, in which all errors are quite satisfactory. These results strongly confirm the performance of the adaptive PINNs, which can be used as a competitive alternative for solving dynamic thermo-mechanical coupling problems in 3D FGMs with complex large-size-ratio structures.

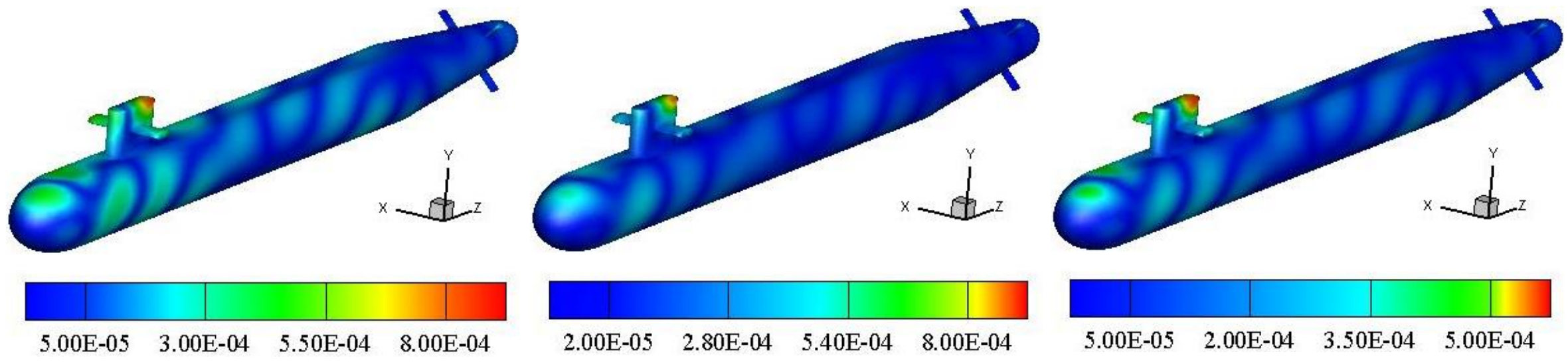

(a) $\sigma_{11}$ (b) $\sigma_{22}$ (c) $\sigma_{33}$

**Fig. 10.** Relative errors of stresses on the submarine surface at $t = 0.9\text{s}$ calculated by the adaptive PINNs.

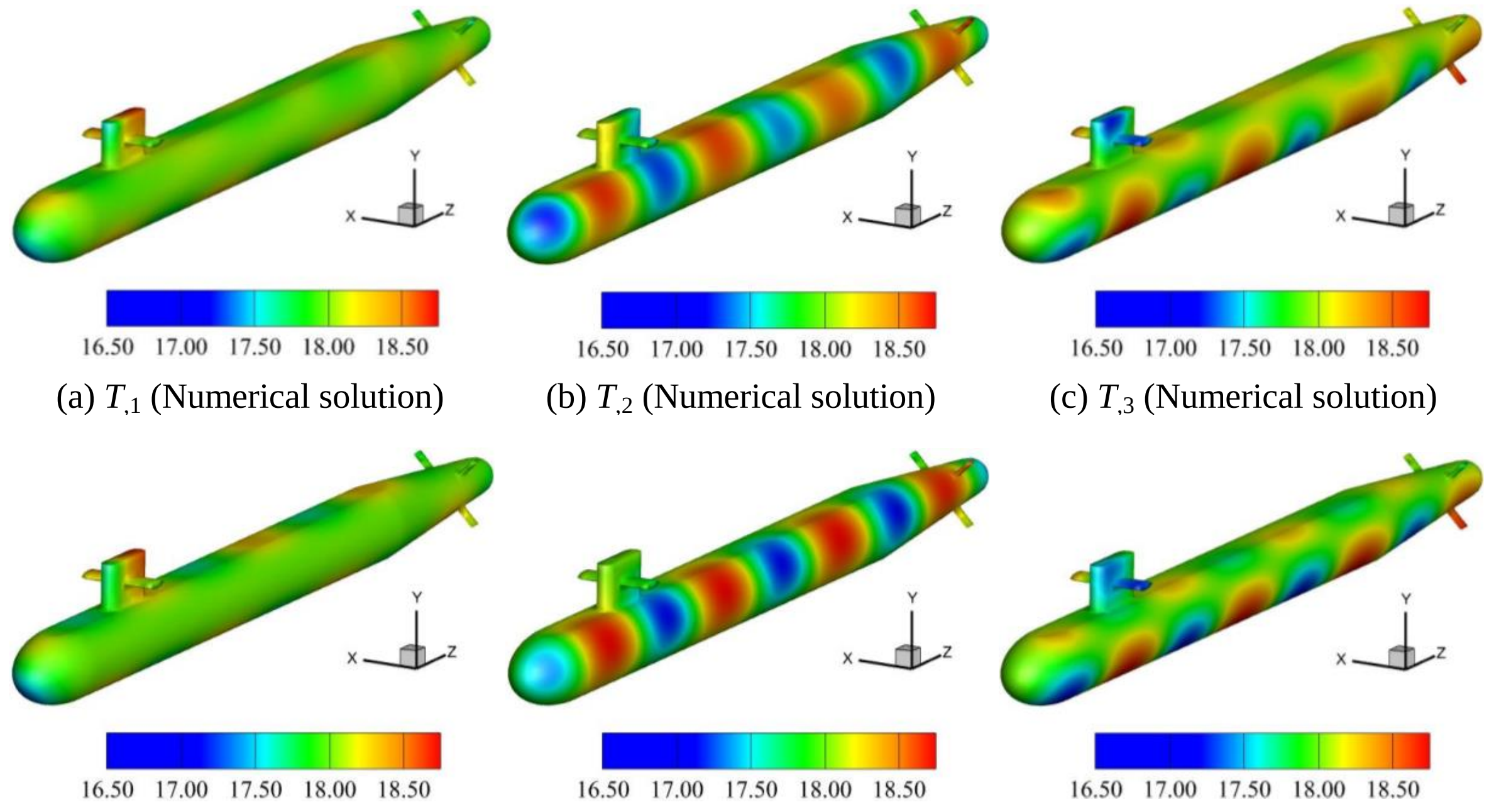

(a) $T_{,1}$ (Numerical solution) (b) $T_{,2}$ (Numerical solution) (c) $T_{,3}$ (Numerical solution)

(d) $T_{,1}$ (Analytical solution) (e) $T_{,2}$ (Analytical solution) (f) $T_{,3}$ (Analytical solution)

**Fig. 11.** Comparison of numerical and analytical heat flux distributions on the submarine surface at $t = 0.5\text{s}$.

**Table 4** Global errors of stresses and heat fluxes evaluated by the adaptive PINNs for the airplane-shaped and the submarine-shaped FGMs.

| Computational domain | Stress | | | Heat flux | | |
|---|---|---|---|---|---|---|
| | $\sigma_{11}$ | $\sigma_{22}$ | $\sigma_{33}$ | $T_{,1}$ | $T_{,2}$ | $T_{,3}$ |
| Airplane | $2.52\times10^{-5}$ | $2.41\times10^{-5}$ | $2.35\times10^{-5}$ | $8.19\times10^{-4}$ | $5.08\times10^{-4}$ | $6.27\times10^{-4}$ |
| Submarine | $1.13\times10^{-4}$ | $1.17\times10^{-4}$ | $1.14\times10^{-4}$ | $4.38\times10^{-3}$ | $4.98\times10^{-3}$ | $3.70\times10^{-3}$ |

## 5. Concluding remarks

In this work, we develop the adaptive PINNs for 3D dynamic thermo-mechanical coupling analysis of FGMs, focusing on the problems with complex large-size-ratio structures that are intractable to conventional approaches. In the proposed PINNs, four neural networks are designed to approximate the displacement components and temperature function, and then these networks are combined together and trained by minimizing a composite loss function based on the coupled PDEs, initial and boundary conditions. In addition, an adaptive loss balancing algorithm is introduced and developed to further improve the performance of PINNs. It is pointed out that the adaptive PINNs are effectively implemented by utilizing the existing optimization and deep learning toolboxes in MATLAB.

Several benchmark examples are provided to examine the validity and feasibility of the adaptive PINNs. It is concluded that the adaptive PINNs are effective and reliable for coating structures with large size ratios up to $10^9$, as well as for complex large-size-ratio structures such as the electrostatic comb, the airplane and the submarine, and can be viewed as a potential alternative for resolving dynamic thermo-mechanical coupling problems in 3D FGMs with complex large-size-ratio geometries. The proposed PINNs has the potential to be extended to other issues, e.g., the problems with large-size-ratio structures and cracks, the thermo-mechanical-chemical coupling analysis [41], the magneto-electro-thermo-mechanical coupled problems [42], etc. Furthermore, some methodologies, such as multi-GPU computing [37], importance sampling method [39], can be considered to further improve the performance of the adaptive PINNs, especially the computational efficiency.

## Declaration of Competing Interest

The authors declare that they have no known competing financial interests or personal relationships that could have appeared to influence the work reported in this paper.

## References

[1] M. Koizumi, FGM activities in Japan, Compos. B. Eng., 28 (1997) 1-4.
[2] B. Saleh, J. Jiang, R. Fathi, T. Al-hababi, Q. Xu, L. Wang, D. Song, A. Ma, 30 Years of functionally graded materials: An overview of manufacturing methods, Applications and Future Challenges, Compos. B. Eng., 201 (2020) 108376.
[3] M.K. Hossain, M. Rubel, M.A. Akbar, M.H. Ahmed, N. Haque, M.F. Rahman, J. Hossain, K.M. Hossain, A review on recent applications and future prospects of rare earth oxides in corrosion and thermal barrier coatings, catalysts, tribological, and environmental sectors, Ceram. Int., (2022).
[4] L.-L. Ke, Y.-S. Wang, Two-dimensional contact mechanics of functionally graded materials with arbitrary spatial variations of material properties, Int. J. Solids Struct., 43 (2006) 5779-5798.
[5] D. He, D. Huang, L. Wu, Y. Xu, Investigation on thermal failure of functionally graded materials using fully coupled thermo-mechanical peridynamics, Compos. Struct., 305 (2023) 116454.
[6] L.D. Bobbio, R.A. Otis, J.P. Borgonia, R.P. Dillon, A.A. Shapiro, Z.-K. Liu, A.M. Beese, Additive manufacturing of a functionally graded material from Ti-6Al-4V to Invar: Experimental characterization and thermodynamic calculations, Acta Mater., 127 (2017) 133-142.
[7] Y. Gu, W. Qu, W. Chen, L. Song, C. Zhang, The generalized finite difference method for long-time dynamic modeling of three-dimensional coupled thermoelasticity problems, J. Comput. Phys., 384 (2019) 42-59.
[8] M. Iqbal, C. Birk, E. Ooi, A. Pramod, S. Natarajan, H. Gravenkamp, C. Song, Thermoelastic fracture analysis of functionally graded materials using the scaled boundary finite element method, Eng. Fract. Mech., 264 (2022) 108305.
[9] M. Malek, N. Izem, M. Seaid, A three-dimensional enriched finite element method for nonlinear transient heat transfer in functionally graded materials, Int. J. Heat Mass Transfer, 155 (2020) 119804.
[10] A. Sutradhar, G.H. Paulino, The simple boundary element method for transient heat conduction in

functionally graded materials, Comput. Methods Appl. Mech. Engrg., 193 (2004) 4511-4539.
[11] H. Wang, Q.-H. Qin, Meshless approach for thermo-mechanical analysis of functionally graded materials, Eng. Anal. Boundary Elem., 32 (2008) 704-712.
[12] L. Qiu, J. Lin, F.J. Wang, Q.-H. Qin, C.-S. Liu, A homogenization function method for inverse heat source problems in 3D functionally graded materials, Appl. Math. Model., 91 (2021) 923-933.
[13] F. Wang, Y. Gu, W. Qu, C. Zhang, Localized boundary knot method and its application to large-scale acoustic problems, Comput. Methods Appl. Mech. Engrg., 361 (2020) 112729.
[14] C.-S. Liu, L. Qiu, Solving nonlinear elliptic inverse source, coefficient and conductivity problems by the methods with bases satisfying the boundary conditions automatically, J. Sci. Comput., 95 (2023) 42.
[15] H.H. Bennani, J. Takadoum, Finite element model of elastic stresses in thin coatings submitted to applied forces, Surf. Coat. Technol., 111 (1999) 80-85.
[16] L. Qiu, F.J. Wang, J. Lin, A meshless singular boundary method for transient heat conduction problems in layered materials, Comput. Math. Appl., 78 (2019) 3544-3562.
[17] Y. Gu, L. Sun, Electroelastic analysis of two-dimensional ultrathin layered piezoelectric films by an advanced boundary element method, Int. J. Numer. Methods Eng., 122 (2021) 2653-2671.
[18] A. Karageorghis, D. Lesnic, L. Marin, The method of fundamental solutions for the detection of rigid inclusions and cavities in plane linear elastic bodies, Comput. Struct., 106 (2012) 176-188.
[19] L. Qiu, F.J. Wang, J. Lin, Q.-H. Qin, Q.H. Zhao, A novel combined space-time algorithm for transient heat conduction problems with heat sources in complex geometry, Comput. Struct., 247 (2021) 106495.
[20] Q. Xi, Z.-J. Fu, T. Rabczuk, An efficient boundary collocation scheme for transient thermal analysis in large-size-ratio functionally graded materials under heat source load, Comput. Mech., 64 (2019) 1221-1235.
[21] Y. Gu, W. Chen, B. Zhang, Stress analysis for two-dimensional thin structural problems using the meshless singular boundary method, Eng. Anal. Boundary Elem., 59 (2015) 1-7.
[22] Z. Fu, Q. Xi, Y. Gu, J. Li, W. Qu, L. Sun, X. Wei, F. Wang, J. Lin, W. Li, Singular boundary method: A review and computer implementation aspects, Eng. Anal. Boundary Elem., 147 (2023) 231-266.
[23] M. Raissi, P. Perdikaris, G.E. Karniadakis, Physics-informed neural networks: A deep learning framework for solving forward and inverse problems involving nonlinear partial differential equations, J. Comput. Phys., 378 (2019) 686-707.
[24] H. Wessels, C. Weißenfels, P. Wriggers, The neural particle method-an updated Lagrangian physics informed neural network for computational fluid dynamics, Comput. Methods Appl. Mech. Engrg., 368 (2020) 113127.
[25] M. Raissi, A. Yazdani, G.E. Karniadakis, Hidden fluid mechanics: Learning velocity and pressure fields from flow visualizations, Science, 367 (2020) 1026-1030.
[26] S. Goswami, C. Anitescu, S. Chakraborty, T. Rabczuk, Transfer learning enhanced physics informed neural network for phase-field modeling of fracture, Theor. Appl. Fract. Mech., 106 (2020) 102447.
[27] A. Harandi, A. Moeineddin, M. Kaliske, S. Reese, S. Rezaei, Mixed formulation of physics-informed neural networks for thermo-mechanically coupled systems and heterogeneous domains, arXiv preprint arXiv:2302.04954, (2023).
[28] Z. Fang, J. Zhan, Deep physical informed neural networks for metamaterial design, IEEE Access, 8 (2019) 24506-24513.

[29] G. Wu, F. Wang, L. Qiu, Physics-informed neural network for solving Hausdorff derivative Poisson equations, Fractals, (2023).

[30] G. Kissas, Y. Yang, E. Hwuang, W.R. Witschey, J.A. Detre, P. Perdikaris, Machine learning in cardiovascular flows modeling: Predicting arterial blood pressure from non-invasive 4D flow MRI data using physics-informed neural networks, Comput. Methods Appl. Mech. Engrg., 358 (2020) 112623.

[31] G.S. Misyris, A. Venzke, S. Chatzivasileiadis, Physics-informed neural networks for power systems, in: 2020 IEEE Power & Energy Society General Meeting (PESGM), IEEE, 2020, pp. 1-5.

[32] R.G. Nascimento, M. Corbetta, C.S. Kulkarni, F.A. Viana, Hybrid physics-informed neural networks for lithium-ion battery modeling and prognosis, J. Power Sources, 513 (2021) 230526.

[33] A.D. Jagtap, E. Kharazmi, G.E. Karniadakis, Conservative physics-informed neural networks on discrete domains for conservation laws: Applications to forward and inverse problems, Comput. Methods Appl. Mech. Engrg., 365 (2020) 113028.

[34] G. Pang, L. Lu, G.E. Karniadakis, fPINNs: Fractional physics-informed neural networks, SIAM J. Sci. Comput., 41 (2019) A2603-A2626.

[35] K. Linka, A. Schäfer, X. Meng, Z. Zou, G.E. Karniadakis, E. Kuhl, Bayesian Physics Informed Neural Networks for real-world nonlinear dynamical systems, Comput. Methods Appl. Mech. Engrg., 402 (2022) 115346.

[36] E. Kharazmi, Z. Zhang, G.E. Karniadakis, hp-VPINNs: Variational physics-informed neural networks with domain decomposition, Comput. Methods Appl. Mech. Engrg., 374 (2021) 113547.

[37] K. Shukla, A.D. Jagtap, G.E. Karniadakis, Parallel physics-informed neural networks via domain decomposition, J. Comput. Phys., 447 (2021) 110683.

[38] L. McClenny, U. Braga-Neto, Self-adaptive physics-informed neural networks using a soft attention mechanism, arXiv preprint arXiv:2009.04544, (2020).

[39] M.A. Nabian, R.J. Gladstone, H. Meidani, Efficient training of physics‐informed neural networks via importance sampling, Comput.-Aided Civ. Infrastruct. Eng., 36 (2021) 962-977.

[40] S. Wang, Y. Teng, P. Perdikaris, Understanding and mitigating gradient flow pathologies in physics-informed neural networks, SIAM J. Sci. Comput., 43 (2021) A3055-A3081.

[41] W. Shan, D. Li, Thermo-mechanic-chemical coupling fracture analysis for thermal barrier coating based on extended layerwise method, Surf. Coat. Technol., 405 (2021) 126520.

[42] Y. Liu, Z. Qin, F. Chu, Investigation of magneto-electro-thermo-mechanical loads on nonlinear forced vibrations of composite cylindrical shells, Commun. Nonlinear Sci. Numer. Simul., 107 (2022) 106146.